\documentstyle[aps,psfig,multicol]{revtex}
\newcommand{\ignore}[1]{}
\begin{document}

\title{Statistics of branched flow in a weak correlated random potential}
\author{Lev Kaplan}
\address{Institute for Nuclear Theory,
 University of Washington, Seattle, Washington 98195}
\maketitle
\begin{abstract}
Recent images of electron flow through a two-dimensional electron gas (2DEG)
device show branching behavior that is reproduced in numerical simulations of
motion in a correlated random potential [3]. We show how such branching
naturally arises from caustics in the classical flow and find a simple scaling
behavior of the branching under variation of the random potential strength.
Analytic results describing the statistical properties of the branching are
confirmed by classical and quantum numerical tests.
\end{abstract}

\begin{multicols}{2}

During the past decade, major advances in the fabrication and study of
mesoscopic (or nanoscale) structures have led to the observation of many novel
phenomena, and a fruitful interaction has resulted with simultaneous
theoretical progress in the areas of random matrix theory, quantum chaos, and
disordered systems. Examples of phenomena that have received much attention
include universal conductance fluctuations and weak localization in open
quantum dots, and the statistics of conductance peak spacings and heights in
the Coulomb blockade regime~\cite{alhassid}.

More recently, scanning probe microscope technology has allowed for the imaging
of current flowing in a 2DEG device~\cite{imaging}. After the electron current
passes  through a narrow quantum point contact (QPC), the flow shows a striking
branch-like behavior~\cite{branch}.  The qualitative features of the observed
branching pattern are well reproduced in numerical simulations of quantum wave
evolution through a correlated random potential, when the rms potential height
and the correlation length of the potential are fixed at their experimentally
measured values.  Perhaps more surprisingly, the same branches are observed in
the corresponding classical simulation, indicating that the experimental
phenomenon should have a classical explanation. However, the simulations show
that the branches {\it do not} correspond to valleys in the random potential.
Instead, it was suggested that branches may arise from caustics in the
classical flow. 

We emphasize that the problem of wave flow through a correlated random
potential is of general interest and not restricted to the mesoscopic context.
For example, similar effects of caustics have been discussed recently in the
context of long-range sound propagation through the ocean~\cite{tomsovic}. We
also note that the influence of caustics on waves in random media has been
studied as far back as 1977 by Berry~\cite{berry}, who analyzed the moments of
the intensity distribution.  The related problem of the effect of orbit
bifurcations on chaotic wave functions is also a research area of continuing
interest~\cite{keating}. Accordingly, our goal here is not only to make
contact with one particular experimental observation but also to
make progress more generally in exploring this regime of classical and quantum 
chaotic behavior.

In the present work, we extend the previous qualitative understanding of
branching behavior, arriving at analytical results to be compared with numerics
and ultimately with experiment. We focus on propagation in a two-dimensional
potential, though the methods apply as well to the three-dimensional case
(where branches are replaced by caustic surfaces) and to related problems such
as light propagation through a medium with varying refractive index.  We
concentrate throughout on generic patterns and not on rare events that would
occur only in a few realizations of the random ensemble.

Our starting point is a Gaussian-distributed random potential $V(x,y)$ with a
Gaussian correlation function,
\begin{eqnarray}
 \overline{ V(\vec r)} &=& 0 \nonumber \\
 \overline{ V(\vec r)V(\vec{r'}) } &=& v_0^2
 e^{-|\vec r - \vec{ r'}|^2/\xi^2} \,.
\label{vstat}
\end{eqnarray}
Without loss of generality, we will use units in which the correlation scale
$\xi=1$; similarly the mass and the initial momentum, taken to be in the
$x-$direction, are set to unity.  To prevent backscattering and localization on
a scale of order $\xi$, we must require the random potential to be weak
relative to the initial kinetic energy, $v_0 \ll E_F=1/2$. In the
experiment~\cite{branch}, $v_0/E_F\approx 0.08$, and all our subsequent
analytical expressions are computed to leading order in $v_0$; however, the
results turn out to be qualitatively and even in some respects quantitatively
valid for $v_0/E_F=0.3$ and higher. In this approximation, then, the motion is
unidirectional: $p_x(t)=p_x(0)\equiv 1$, $x(t)=t$, and
\begin{equation}
 dy(t) / dt =p_y(t)\,, \;\;\;\;\; dp_y(t) / dt =
 -{\partial V(t,y)/\partial y} \,.
\end{equation}
In other words, the two-dimensional dynamics reduces to one-dimensional
evolution in the transverse dimension $y$, under the influence of an
effectively time-varying random potential~\cite{cons}. For numerical purposes,
the above classical or quantum $y$-evolution may be discretized on a scale
$\Delta t = \Delta x \ll 1$. The initial condition is zero transverse momentum;
classically it is the manifold $p_y=0$ in the $y-p_y$ phase space.

Stretching and folding of this initial manifold under classical evolution in
$t$ (or $x$) produces simple caustics, or folds, where  $\delta p_y(t)/ \delta
y(0) =\infty$ and the classical intensity in position space diverges~\cite{ho}.
By calculating how long it takes for a trajectory to travel a distance of one
correlation length in the transverse $y-$direction, we may easily check that
this stretching and folding occurs in the classical dynamics on the
characteristic time scale
\begin{equation}
\label{t0}
 t_0 \equiv v_0^{-2/3} \gg 1\,,
\end{equation}
i.e., when the potential is weak, a trajectory must pass over many correlation
lengths of the potential in the longitudinal ($x$) direction before caustics
arise. In the following, we eliminate the explicit dependence of branching
statistics on the random potential strength $v_0$ by expressing all results in
terms of the scaled time $t/t_0$.

For $t/t_0 > 1$, the transverse stretching factor defined as $s\equiv\delta
y(t)/ \delta y(0)$ develops a log-normal distribution,
\begin{equation}
 \ln P(s) \approx -{(\ln s - \alpha t/t_0)^2 \over \beta t/t_0} \,,
\label{ps}
\end{equation}
where the exponent $\alpha$ is a dimensionless Lyapunov exponent, $
\overline{\ln s} = \alpha t/t_0$, while $\beta$ characterizes the inhomogeneity
of the stretching, $\overline{(\delta \ln s)^2} =(\beta/2) t/t_0$. The average
amount of stretching in the $y-$direction of an infinitesimal piece of the
initial manifold grows exponentially with time as
\begin{equation}
\label{meanstretch}
 \overline{s} = \overline{ \delta y(t)/\delta y(0)}
 \sim e^{(\alpha+\beta/4)t/t_0}\,.
\end{equation}
Since folds in the classical manifold typically develop when $\delta y \sim \xi
=1$, the number of caustics also grows
exponentially with time as \begin{equation}
\label{ncaus}
 N^{\rm caus} \sim We^{(\alpha+\beta/4)t/t_0}\,,
\end{equation}
where $W$ is the length in the $y-$direction of the initial manifold, i.e., the
transverse width of the system in units of the correlation length. The scaling
(\ref{ncaus}) has been checked numerically using several values of the random
potential strength $v_0$.

Because the classical density of trajectories diverges at a caustic, we must
perform smoothing of intensity on some scale $b \ll 1$ in the transverse
$y-$direction in order to obtain a well-defined branch height. In quantum
mechanics, such smoothing is of course taken care of automatically by the
uncertainty principle, since caustics cannot be resolved when the phase space
area enclosed by the fold is below $\hbar$, and thus $b_{\rm eff} \sim
\hbar^{2/3}$ (see below). We also adopt the convention that the initial
intensity (classical density of trajectories or quantum wave function density)
is normalized to unity.

When a caustic forms at time $t_{\rm form}$ and transverse position $y$, the
classical $b-$smoothed intensity $I_b(t_{\rm form},y)$ of the resulting branch
scales as 
\begin{equation}
I_b \sim b^{-1/2}[s(0,t_{\rm form})]^{-1} \,,
\label{ib}
\end{equation}
where $s(0,t_{\rm form})$ is of course the stretching factor between time $0$
and time $t_{\rm form}$ of the piece of the manifold in which the fold occurs.
Away from caustics the intensity scales simply as $I_b \sim
[s(0,t)]^{-1}$. The extra factor of $b^{-1/2}$ in Eq.~(\ref{ib}) ensures that
for small $b$ (corresponding to the semiclassical or ray limit of the original
quantum or wave problem), caustics always dominate the tail of the intensity
distribution, as we have confirmed numerically. 

Using the typical behavior of the stretching factor $s$ (Eq.~(\ref{ps})), we
now immediately see from Eq.~(\ref{ib}) that for $\alpha^{-1} < t_{\rm
form}/t_0 < {1 \over 2} \alpha^{-1}\ln b^{-1}$, every caustic leads to a
visible branch, with intensity $I_b \gg 1$.  Once a given branch forms, its
intensity immediately begins to decay (though at a slower rate) due to further
stretching: $I_b(t,y) \sim b^{-1/2}[s(0,t_{\rm form})]^{-1}[s(t_{\rm
form},t)]^{-1/2}$, eventually vanishing into the background on a logarithmic
time scale $t/t_0 \sim \alpha^{-1}\ln b^{-1} \gg 1$. The number of visible
branches reaches its peak value of $N^{\rm branch} \sim Wb^{-1/2}$ at the time
scale $t/t_0 \approx {1 \over 2}\alpha^{-1} \ln b^{-1}$.

At longer times, $t/t_0 > {1 \over 2} \alpha^{-1}\ln b^{-1}$, it is no longer
true that a typical new caustic results in a visible branch, since the
resulting smoothed intensity is generally below the level ($I_b \approx 1$) of
the random background. Thus, at these longer times branches appear only at
caustics that are formed from pieces of the manifold that have experienced
anomalously little stretching compared with the average behavior.  The number
of such visible branches may be easily estimated by first calculating the
number of caustics coming from those regions of the manifold that have
stretched by no more than some factor $s_0 \ll \exp(\alpha t/t_0)$. This number
is given by the product of the manifold fraction that has stretched by less
than $s_0$ and the number of caustics per unit length resulting from that
stretching:
\begin{equation}
\label{ncaus2}
 \ln { N^{\rm caus}_{s<s_0} \over W} \approx 
 {-(\ln s_0 - \alpha t/t_0)^2 \over \beta
 t/t_0} + \ln s_0 \,.
\end{equation}
Inserting the maximal stretching factor $s_0 = e^{(const/2)} b^{-1/2}$ that
will still allow a newly-formed caustic to produce a visible branch, we obtain
for $t/t_0>{1 \over 2} \alpha^{-1}\ln b^{-1}$
\begin{equation}
 \ln {N^{\rm branch} \over W} \approx \left({1 \over 2}+{\alpha \over
 \beta}\right) \left(\ln b^{-1} + const\right) -
 {\alpha^2 \over \beta} {t\over t_0} +O \left({t_0 \over t}\right) \,.
\label{nbranch}
\end{equation}
Thus, after a rapid initial climb, the number of branches falls off
exponentially with scaled time $t/t_0$, at the rate $\alpha^2/\beta$. The
inherent ambiguity in defining how high a branch must be above the background
to be visible can be absorbed into a multiplicative constant in $N^{\rm
branch}$ and does not affect the falloff {\it rate}.

\begin{figure}[h] \begin{center} \leavevmode \parbox{0.5\textwidth}
{
\psfig{file=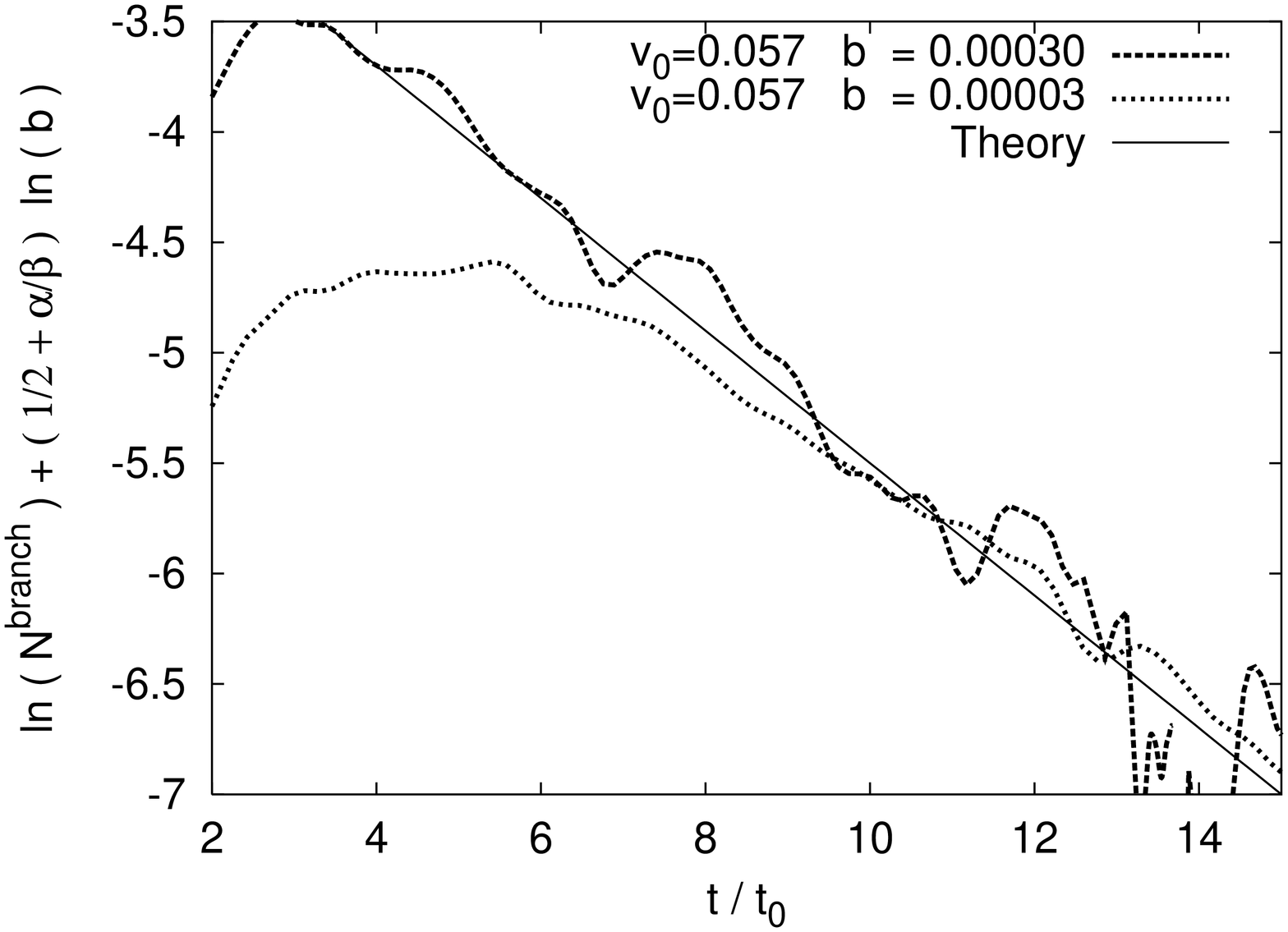,width=0.45\textwidth,angle=270}
}
\end{center}
\protect\caption{
Number of visible branches as a function of dimensionless scaled time $t/t_0$,
where the characteristic time scale $t_0$ depends on the strength $v_0$ of the
random potential, Eq.~(\ref{t0}).  A visible branch is defined as an intensity
enhancement of at least a factor of $10$ above the background.  The two dashed
curves correspond to different values of the smoothing length $b$.  The solid
line represents the predicted exponential falloff at long times
(Eq.~(\ref{nbranch})), with slope $\alpha^2/\beta=0.30$ ($\alpha=0.65$,
$\beta=1.42$).
}
\label{fignbr}
\end{figure}

Numerical results are presented in Fig.~\ref{fignbr}. In this figure and
throughout, a single realization of the random ensemble is used for each data
curve; no ensemble averaging is necessary to obtain convincing statistics.

\begin{figure}[h] \begin{center} \leavevmode \parbox{0.5\textwidth}
{
\psfig{file=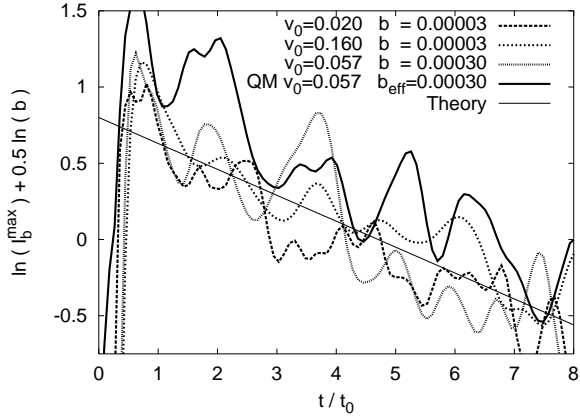,width=0.45\textwidth,angle=270}
}
\end{center}
\protect\caption{Exponential falloff of maximum branch intensity $I_b^{\rm
max}$ with scaled time $t/t_0$.  The three dashed curves correspond to
classical simulations for several values of $v_0$ and smoothing scale $b$, and
the solid curve is a quantum calculation.  The power-law dependence of
$I_b^{\rm max}$ on $b$ confirms that the peaks arise from simple caustics
(folds). Finally, the straight line corresponds to the theoretical prediction
of Eq.~(\ref{imax}), with analytically predicted exponent $\gamma=0.17$
($\alpha=0.65$, $\beta=1.42$).}
\label{figimax}
\end{figure}

In the experimental electron flow images~\cite{branch}, the few longest and
most intense branches are the most visually striking.  Let us therefore focus
on the maximum branch intensity $I_b^{\rm max}$ as a function of scaled time
$t/t_0$. This from Eq.~(\ref{ib}) is simply $b^{-1/2}$ divided by the {\it
minimum} stretching factor, which is immediately found by setting $N^{\rm
caus}_{s<s_0}=1$ in Eq.~(\ref{ncaus2}) and solving for $\ln s_0$. We obtain
\begin{eqnarray}
& \ln & I_b^{\rm max}  \approx  {1 \over 2} \ln b^{-1} -
\gamma {t\over t_0} \nonumber \\
\gamma & = & \alpha - {\beta \over 2}\left(\sqrt{1+{4 \alpha \over
\beta}}-1\right) \,.
\label{imax}
\end{eqnarray}
This predicted exponential falloff of $I_b^{\rm max}$ with $t/t_0$ is confirmed
in Fig.~\ref{figimax}.  We note the correct scaling behavior of the maximum
intensity for different values of the potential strength $v_0$ and different
smoothing lengths $b$, as well as good agreement between the theoretically
predicted and numerically observed numerical exponent $\gamma$.  The behavior
$I_b^{\rm max} \sim b^{-1/2}$ confirms that the maximum intensities indeed
always come from caustics, which we have also confirmed by direct examination.

Due to intensity fluctuations within a given branch, particularly in the
quantum case, the fraction of space covered by branches, $f^{\rm branch}$, is
in practice a more robust measure than the branch number $N^{\rm branch}$
(Eq.~(\ref{nbranch})). Defining $f^{\rm branch}$ as the fraction of intensities
that exceed some arbitrary cutoff $I^{\rm cutoff}$, and combining the result of
Eq.~(\ref{imax}) with the fact that intensity falls off as the inverse square
root of the distance as we move away from a caustic, one straightforwardly
obtains (for $\beta \ge 2 \alpha$)
\begin{equation}
\ln f^{\rm branch} \approx  -2\gamma {t \over t_0} -2 \ln I^{\rm cutoff}
\label{fraction}
\end{equation}

\begin{figure}[h] \begin{center} \leavevmode \parbox{0.5\textwidth}
{
\psfig{file=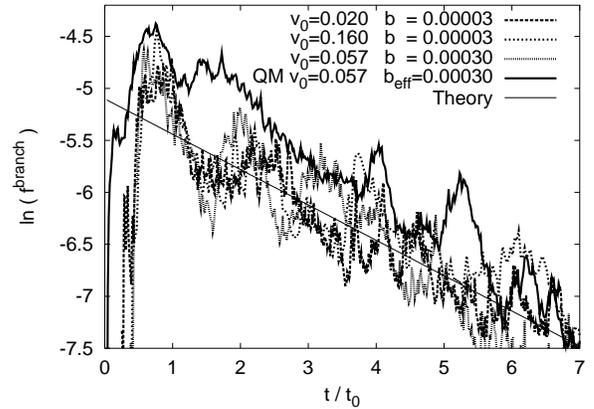,width=0.45\textwidth,angle=270}
}
\end{center}
\protect\caption{Fraction of space $f^{\rm branch}$ covered by branches taller
than the arbitrary cutoff $I^{\rm cutoff}=10$, as a function of scaled time
$t/t_0$.  Again, correct scaling is observed under changes of the potential
strength $v_0$. The slope of the solid line is the analytical prediction
$2\gamma=0.34$ of Eq.~(\ref{fraction}).}
\label{figbig}
\end{figure}

\begin{figure}[h] \begin{center} \leavevmode \parbox{0.5\textwidth}
{
\psfig{file=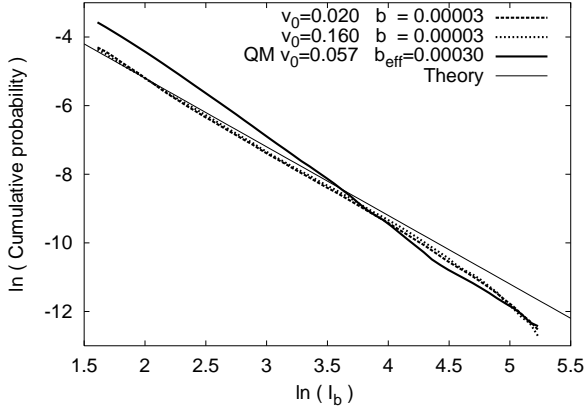,width=0.45\textwidth,angle=270}
}
\end{center}
\protect\caption{Cumulative probability distribution $\int_{I_b}^\infty dI_b'
\; P(I_b')$ of intensities $I_b$ at a fixed scaled time $t/t_0=2.2$.  The slope
$-2$ of the solid line is the theoretical prediction of Eq.~(\ref{cumul}). Note
that the power-law behavior breaks down as we approach the maximum intensity
from Eq.~(\ref{imax}), $I_b^{\rm max}\approx e^{4.5}$ for $b=0.0003$ and
$I_b^{\rm max}\approx e^{5.5}$ for $b=0.00003$.
}
\label{figdistr}
\end{figure}

Eq.~(\ref{fraction}) predicts that irrespectively of our definition of a branch
(i.e., choice of $I^{\rm cutoff}$), the spatial area covered by visible
branches falls off exponentially with distance from the starting point, and the
exponent is given by $2\gamma$. This behavior is confirmed in Fig.~\ref{figbig}
for several sets of parameters. We note in particular that to leading order the
result does not depend on the smoothing scale $b$, and correspondingly on the
ratio of wavelength to correlation length in the quantum or wave problem.
Eq.~(\ref{fraction}) may alternatively be interpreted as giving the cumulative
distribution of intensities at a fixed scaled time $t/t_0$:
\begin{equation}
\int_{I_b}^\infty dI_b' \; P(I'_b)  \sim \exp\left(-2\gamma{t \over t_0}\right)
I_b^{-2}\,, 
\label{cumul}
\end{equation}
i.e., $P(I_b) \sim I_b^{-3}$. This power-law behavior remains valid until
we reach the maximum intensity as given by Eq.~(\ref{imax}). The $I_b^{-3}$
falloff means that tall branches do not affect the average intensity, but the 
mean squared intensity or inverse participation ratio (IPR) is logarithmically
divergent in the smoothing scale $b$. The predicted power-law intensity
distribution at fixed time is tested numerically in Fig.~\ref{figdistr}.

By explicitly tracking all caustic locations for the classical dynamics in
Fig.~\ref{figdistr}, we have also checked that the fraction of large
intensities $I_b$ that come from branches associated with caustics ranges from
$98.7\%$ for $\ln I_b \ge 2$ to $99.8\%$ for $\ln I_b \ge 4$, while only
$3.8\%$ of space overall is covered by such branches at time $t/t_0 =2.2$.
This again provides confirmation that caustics dominate the tail of the
intensity distribution.

Our classical results for the branching statistics can be directly applied to
the quantum context once we understand the smoothing scale $b_{\rm eff}$
implied by the uncertainty principle in the quantum case.  The parabolic
structure of the classical manifold near a fold at $(y^\ast,p^\ast_y)$ takes
the form $y-y^\ast \sim [(p_y-p_y^\ast)/p_y^{\rm typ}]^2$, where $p_y^{\rm
typ}$ is the typical value of the transverse momentum, $p_y^{\rm typ} \sim v_0
t^{1/2} \sim v_0^{2/3} (t/t_0)^{1/2}$. The area enclosed by the fold then
scales as $(y-y^\ast)(p_y-p_y^\ast) \sim (y-y^\ast)^{3/2} p_y^{\rm typ} \sim
(y-y^\ast)^{3/2} v_0^{2/3}$. Setting this area to $\hbar$ and dropping factors
of order unity, we finally obtain $y-y^\ast \sim \hbar^{2/3} v_0^{-4/9}$, i.e.,
caustics in quantum mechanics do not get resolved below an effective smoothing
scale
\begin{equation}
 b_{\rm eff} = \hbar^{2/3} v_0^{-4/9}\,.
\label{beff}
\end{equation}
We may now simply transcribe our previously obtained classical results,
substituting the effective quantum smoothing scale $b_{\rm eff}$ for the
explicit classical smoothing scale $b$. Data from typical quantum calculations
appear as solid curves in Figs.~\ref{figimax}-\ref{figdistr}.

In summary, we have made progress towards an analytic and quantitative
understanding of branching behavior for flow in a random correlated potential.
The key conditions of applicability of these results are (1) a weak random
potential $v_0/E_F \ll 1$, (2) a focus on distance scales of order the Lyapunov
length, $L_{\rm Lyap} \sim \xi (v_0/E_F)^{-2/3}$, where $\xi$ is the
correlation scale of the potential, and (3) in the quantum case a wavelength no
larger than $\xi$ to allow for a simple semiclassical treatment of the
caustics.  We mention in conclusion that the {\it numerical values} of the
exponents $\alpha$, $\beta$ governing the branching statistics may well depend
on the details of the random potential, in particular on the precise
distribution of potential heights and on the two-point correlation function of
the potential (see Eq.~(\ref{vstat})). The overall formalism, on the other
hand, is more generally applicable provided the log-normal behavior of
Eq.~(\ref{ps}) is satisfied.

\section*{Acknowledgments}
Helpful discussions with S. E. J. Shaw and E. J. Heller are gratefully
acknowledged. This work was supported by the U.S. Department of Energy, under
Grant DE-FG03-00ER41132.

\end{multicols}


\begin{references}
\bibitem{cons} The approximation is self-consistent because $p_y$ and $p_x-1$
 remain small ($O(v_0^{2/3})$) over time scales of interest, as we will see
 below.
\bibitem{ho} Higher-order caustics may also occur, but these have codimension
 greater than 1 and produce only isolated intensity peaks, not the extended
 branches that we focus on here.
\end{references}
\end{document}